\documentclass[structabstract]{aa}
\usepackage{graphicx}
\usepackage{txfonts}
\usepackage{natbib}

\bibpunct{(}{)}{;}{a}{}{,} 

\begin{document}

\title{Evolution of the Solar Magnetic Flux on time scales of years to millenia} 

\titlerunning{Evolution of the Solar Magnetic Flux}

\author{Luis Eduardo A. Vieira\inst{1}
  \and Sami K. Solanki\inst{1,2}} 

\offprints{L.E.A. Vieira, \email{vieira@mps.mpg.de}}

\institute{Max-Planck-Institut f\"ur Sonnensystemforschung, 
Max-Planck-Str. 2, 37191 Katlenburg-Lindau, Germany \and School of Space Research, Kyung Hee University, Yongin, Gyeonggi, 446-701, Korea} 

\date{Received 11 September 2009 / Accepted 11 November 2009}

\abstract {} {We improve the description of the evolution of the Sun's open and total magnetic flux on time scales of years to millenia.}
{In the model employed here the evolution of the solar total and open magnetic flux is computed from the flux emerging at the solar surface in the form of bipolar magnetic features, which is related to the sunspot number cycle parameters and can be estimated from historical records. Compared to earlier versions of the model in addition to the long-lived open flux, now also a more rapidly decaying component of the open flux is considered. The model parameters are constrained by comparing its output with observations of the total surface magnetic flux and with a reconstruction of the open magnetic flux based on the geomagnetic indexes. A method to compute the Sun's total magnetic flux and the sunspot number during the Holocene, starting from the open flux obtained from cosmogenic isotopes records, is also presented.}
{By considering separately a rapdly evolving and a slowly evolving component of the open flux the model reproduces the Sun's open flux, as reconstructed based on the aa-index, much better and a reasonable description of the radial component of interplanetary magnetic field data are obtained. The greatest improvement is in the reproduction of the cyclic variation of the open flux, including the amplitudes of individual cycles. Furthermore, we found that approximately 25\% of the modeled open flux values since the end of the Maunder Minimum are lower than the averaged value over 2008, i.e. during the current low minimum. The same proportion is observed in reconstructions of the open flux during the Holocene based on cosmogenic isotopes, which suggests that the present solar minimum conditions are below average, but not exceptional in terms of the heliospheric magnetic flux.}{}


\maketitle 


\section{Introduction}

The variability of the magnetic field has a strong influence on the dynamics of the outer layers of the Sun. Thus, the 11-year cyclic variability of the magnetic field is registered by several solar parameters such as the sunspot number and area, the rate at which flares and coronal mass ejections occur, the flux of solar X-rays, radio waves and solar energetic particles, as well as the total and spectral solar irradiance. The variation of the total magnetic flux and the surface distribution of the field also influences the open magnetic flux and hence the heliospheric magnetic field. 

Our knowledge of the evolution of the Sun's magnetic flux on longer time scales is limited by the availability of continuous and reliable observations of the solar magnetic field. Most of these observations are available for just a few decades, e.g. since the beginning of the space age in the case of the open magnetic flux. On longer time scales, the solar magnetic flux must be reconstructed or computed from proxies. Thus, the heliospheric flux (i.e. the solar open flux)  is reconstructed based on the geomagnetic aa-index from 1868 to the present \citep{lockwood1999, rouillard2007}. The open and total magnetic flux since roughly 1610 can be computed from the sunspot number \citep{solanki2000, solanki2002}. During this period, the open flux doubled. On millennial time scales, reconstructions of the solar open flux and sunspot number were obtained based on cosmogenic isotopes such as \element[ ][14]{C} and \element[ ][10]{Be} \citep{solanki2004, usoskin2002, usoskin2003, usoskin2004, usoskin2007}.

The evaluation of the solar surface magnetic flux budget requires continuous and full-disk magnetograms, which are available for just for the last few solar cycles. Even for this period the total amount of magnetic flux emerging in small bipolar magnetic regions is uncertain due to the limited spatial resolution of the data \citep{krivova2004}. In order to assess the solar magnetic flux budget on longer time scales semi-empirical models have been used \citep{krivova2007, solanki2000, solanki2002}, as well as flux transport computations \citep{baumann2004, mackay2002, schussler2006, wang2002, wang2005}. The former are based on the sunspot record and attempt to reconstruct the evolution of the total magnetic flux and the flux emerging in large and small bipolar regions as well as the evolution of large unipolar regions that give rise to the solar open flux. These models are validated by comparing them with the record of the total magnetic flux obtained from magnetograms as well as with a longer record of the Sun's open magnetic flux deduced from the aa index, a measure of the variability of the Earth's magnetic field produced by its interaction with the variable interplanetary field, i.e. the Sun's open flux \citep{lockwood1999}. 

\begin{figure*}
\centering
\includegraphics[width=13cm,clip=true, viewport=0.1cm 0.1cm 19cm 25cm]{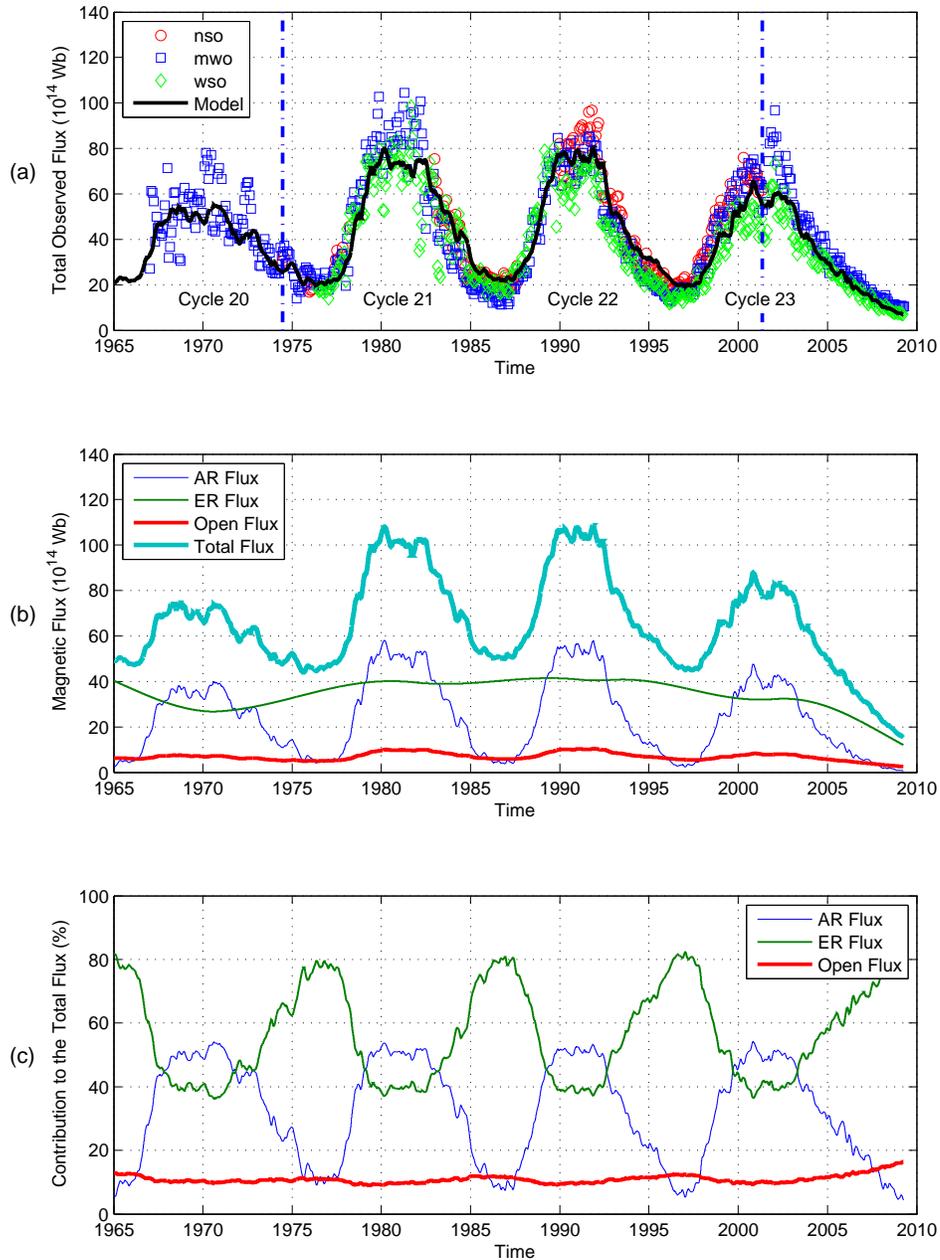}
\caption{(a) Comparison of the observed (symbols) and modeled (solid line) total magnetic flux. Each data point is an integral over a synoptic chart for one Carrington rotation. Different symbols are used for different data sets: circles represent the KP NSO data, squares MWO data and diamonds WSO data. For the modeled flux the value ($\phi_{act}+0.3\phi_{eph}+\phi_{open}$) is given (see text). The dashed blue lines bound the period used for the optimization of the parameters of the model between 1974 and 2002. (b) Reconstructed magnetic flux in AR (blue line), ER (green line), open (Red Line) and total flux (cyan line). For the total flux here the value ($\phi_{act}+\phi_{eph}+\phi_{open}$) is plotted. (c) Fractional contribution of AR, ER and open flux to the total Flux.}
\label{FigTotalFlux}%
\end{figure*}

The models of \cite{solanki2000, solanki2002} of the evolution of the solar magnetic flux reproduce the long-term variation of the solar open flux reasonably well, but give a much smoother variation over the solar cycle than the observations. In addition, the computed open flux lags the observations by roughly 2-3 years. These shortcomings are the result of the simplicity of the model, which, in order to explain the long-term trend requires the open flux to have a lifetime of multiple years. Such an extended lifetime is appropriate for the field in the large polar coronal holes, but not for some of the smaller, relatively short-lived low-latitude coronal holes that are the source regions of a significant fractions of the open flux during the high activity phases of solar cycle.

The main motivation of this work is to estimate the evolution of the solar magnetic flux taking into account that in addition to the long-lived open flux there are also more rapidly evolving source regions (coronal holes). These are often associated with active regions or decaying active regions. For the purposes of the model, it is important to distinguish between the slowly and rapidly evolving flux and not between the locations on the solar surface. In particular, there is no one-to-one correspondence of high and low latitude coronal holes to slowly and rapidly evolving flux. 

The paper is structured as follows. In Sect. 2 we describe the solar magnetic flux model, the parameter optimization procedure, and the model parameters. The results are described in Sect. 3. The extension of the model to periods prior to telescopic sunspot observations is presented in Sect. 4. The conclusions are given in Sect. 5. 

\section{Model Description}

\subsection{Approach}

The model presented here is an extension of the one presented by \cite{solanki2002}, which is itself an extension of the work of \cite{solanki2000}. In this model the evolution of the total and open magnetic flux is computed from the flux emerging at the solar surface in form of bipolar magnetic features. On the Sun, the bipolar magnetic regions display a continuous size spectrum \citep{harvey1993}. In the  model the spectrum is divided into two classes according to the size and life-time of the structures, following \cite{harvey1993} and others. Large bipolar structures emerging in the activity belts and living up to several weeks are classified as Active Regions (AR). Small short-lived bipoles, which emerge over a larger range of latitudes, are classified as Ephemeral Regions (ER). The magnetic flux emergence rate in active regions is roughly proportional to the sunspot number, which allows it to be estimated from historical records. The emergence rate of magnetic flux in ephemeral regions is higher than in active regions and their contribution to the total photospheric magnetic flux is significant. In spite of their limited lifetime the number and latitude of ephemeral regions also evolves over a cycle, which is extended with respect to the sunspot cycle, but shows a much smaller contrast between activity maximum and minimum. The literature contains contradictory statements whether there are more or less ephemeral regions at activity maximum \citep{hagenaar2001, harvey1993}. In the model, parameters of the cycle (such as time of maximum, amplitude and length) displayed by the ephemeral regions are assumed to be related to the properties of the corresponding sunspot cycle. 

Part of the magnetic flux that emerges in active and ephemeral regions is dragged outward by the solar wind and reaches far into the heliosphere. It is called the open magnetic flux. As the source of the open magnetic flux is located in often large regions with a dominant magnetic polarity, it can survive on the solar surface for a relatively long time, reaching up to several years. There are, however, also smaller, shorter lived coronal holes often associated with decaying active regions. These lead to a far more rapid variation in the Sun's open magnetic flux, in particular around activity maximum and shortly after it. Hence, some of the flux from active regions does open, but stays open only for a relatively short time \citep{cranmer2002}. \cite{ikhsanov1999} show a histogram of equatorial coronal holes (CH) lifetimes according to which nearly 50\% do not outlive 3 solar rotations, implying a median lifetime of 80-90 days. This may be an upper limit, since they only consider equatorial coronal holes that survive at least 2 solar rotations. We stress, however, that our rapidly and slowly decaying open fluxes cannot be simply associated with low and high latitude coronal holes. An alternative interpretation of the rapidly evolving open flux is that it is additional flux carried into the heliosphere by CMEs before their disconnection from the Sun \citep{crooker2002, luhmann1998, owens2006}. As such, it is not strictly 'open' (in the sense that it does not reach all the way out to the Heliopause), but it does contribute to the interplanetary field and the Sun's magnetic flux at 1 AU. This is a relevant quantity for comparing with the open magnetic flux at 1 AU  reconstructed by \cite{lockwood2009pers}.


In the following, we extend the previous model of \cite{solanki2002} by distinguishing between rapidly decaying open flux, $\phi_{open}^r$, and slowly decaying open flux, $\phi_{open}^s$. A set of four coupled ordinary differential equations then describes the evolution of the four surface magnetic flux components that we consider:

\begin{equation}
\frac{d\phi_{act}}{dt}  =  \epsilon_{act}-\frac{\phi_{act}}{\tau_{act}^0}-\frac{\phi_{act}}{\tau_{act}^1}-\frac{\phi_{act}}{\tau_{act}^2} \,, \label{eq01}
\end{equation}

\begin{equation}
\frac{d\phi_{eph}}{dt}  =  \epsilon_{eph}-\frac{\phi_{eph}}{\tau_{eph}^0}-\frac{\phi_{eph}}{\tau_{eph}^1} \,, \label{eq02}
\end{equation}

\begin{equation}
\frac{d\phi_{open}^r}{dt}=\frac{\phi_{act}}{\tau_{act}^2}-\frac{\phi_{open}^r}{\tau_{open}^r} \,, \label{eq03}
\end{equation}

\begin{equation}
\frac{d\phi_{open}^s}{dt}=\frac{\phi_{act}}{\tau_{act}^1}+\frac{\phi_{eph}}{\tau_{eph}^1}-\frac{\phi_{open}^s}{\tau_{open}^s} \,, \label{eq04}
\end{equation}

\begin{equation}
\phi_{open}=\phi_{open}^r+\phi_{open}^s \,,  \label{eq05}
\end{equation}

\begin{equation}
\phi_{total}=\phi_{act}+\phi_{eph}+\phi_{open} \,, \label{eq06}
\end{equation}
where $\phi_{act}$, $\phi_{eph}$, $\phi_{open}$, and $\phi_{total}$ refer to magnetic flux of AR, ER, open flux, and the total flux. The open flux is the sum of the rapidly ($\phi_{open}^r$) and slowly ($\phi_{open}^s$)  evolving components (Eq. \ref{eq05}). Distinguishing between these two components of the open flux is the main difference to the computations of magnetic flux evolution by \cite{solanki2002} and \cite{krivova2007} and leads to the introduction of one more differential equation. The $\phi_{open}^r$ is fed only by $\phi_{act}$ since it is assumed to reside in small coronal holes located close to ARs. In contrast, $\phi_{open}^s$ obtains contributions from both active regions and ephemeral regions. It resides at least partly in the polar cap coronal holes, but also elsewhere (e.g. in the flux from decaying active regions wandering to the poles).

The time constants  $\tau_{act}^0$, $\tau_{eph}^0$, $\tau_{open}^r$ and $\tau_{open}^s$ are the decay time scales of the AR, ER, rapid and slow components of the open flux, respectively. Following \cite{solanki2000} and others, we assume that the decay process is due to the cancellation with flux of opposite polarity, but do not specify the process beyond giving the decay time. The time constant $\tau_{act}^2$ is the flux transfer time from active regions to the rapidly evolving component of the open flux, while $\tau_{act}^1$ and $\tau_{eph}^1$  are the flux transfer times from AR and ER to the slowly evolving component of the open flux, respectively. 

The input parameters of the model are the flux emergence rates of active and ephemeral regions. Here, the flux emergence rate in active regions is chosen to be linearly proportional to the monthly averaged group sunspot number \citep{hoyt1998}, $R_g$, and is scaled according to the observations of \cite{schrijver1994} for cycle 21. Following \cite{krivova2007}, we define the flux emergence rate of AR as

\begin{equation}
\label{eq07}
\epsilon_{act}=\epsilon^{max,21}  \frac{R_g}{R_g^{max,21}} \,, 
\end{equation}
with  $\epsilon_{act}^{max,21}=2.3 \times 10^{24}$  Mx yr$^{-1}$  and   $R_g^{max,21}=172$. The value of $R_g^{max,21}$ was obtained from 3-month running means of $R_g$. 

Following \cite{solanki2002}, we obtain the total flux emergence rate in ER as a sum over multiple overlapping cycles 

\begin{equation}
\epsilon_{eph}(t) = \sum_{i=1}^{N_{cycles}} \epsilon_{eph}^i(t) \,, \label{eq08}
\end{equation}
where we define $\epsilon_{eph}^i(t)$ as 

\begin{equation}
\epsilon_{eph}^i(t)=\epsilon_{act}^{max,i}  X  g^i(t) \, .\label{eq09}
\end{equation}

Here $X$  is a scaling factor and $g^i(t)$ is a function defined as

\begin{equation}
\label{eq10}
g^i(t) =
\cases{
cos^2\left(\pi( t - t_c^i )/T^i_{eph}\right), &  $-T^i_{eph}/2 \le (t - t_c^i) \le T^i_{eph}/2 ,$ \cr
0, & other wise.\cr
}
\end{equation}
where $t_c^i$ is the time at which cycle $i$ reaches maximum activity and  $T^i_{eph}$   is the length of the ephemeral cycle. Equation (\ref{eq10}) is equivalent to Eq. (6) of \cite{krivova2007}. The length of the ephemeral cycle is related to the length of the activity cycle, $T^i$, in the following way:

\begin{equation}
\label{eq11}
T_{ext}^i=T^i-c_x \,,   
\end{equation}
and
\begin{equation}
T^i_{eph}=T^i+2 T_{ext}^i \label{eq12} \,,
\end{equation}
where $c_x$ is the ER cycle extension parameter and $2T_{ext}^i$ is the extension of the ER cycle $i$ in relation to AR cycle $i$. 

We define the maximum emergence rate at cycle $i$, which appears in Equation (\ref{eq09}) as 

\begin{equation}
\epsilon_{act}^{max,i}=\epsilon_{act}^{max,21} \frac{R_g^{max,i}}{R_g^{max,21}} \,, \label{eq13} 
\end{equation}
where $R_g^{max,i}$ is the maximum sunspot number observed at cycle $i$.

\subsection{Parameter optimization}

The model has 8 free parameters. These are introduced and discussed in Sect. 2.3. Here we describe how the optimization is carried out. The free parameters were adjusted by comparing the model output with observations of the total surface magnetic flux deduced from synoptic charts of the Sun's radial field \citep{arge2002} and with a reconstruction of the open magnetic flux based on the geomagnetic aa-index by \cite{lockwood2009pers}. This is a revised version of the reconstruction due to \cite{lockwood1999}.  This revised open flux includes corrections due to kinematic effects produced by the propagation of CMEs (Lockwood et al. 2009a, b). It covers the period from 1904 to 2008.

The optimization of the model's free parameters is realized using the genetic algorithm PIKAIA described by \cite{charbonneau1995}. The code maximizes a function (fitness function, $f$), defined as 

\begin{equation}
f(parameters)=\frac{1}{{\chi\prime}_{total}^2+{\chi\prime}_{open}^2} \,, \label{eq14}
\end{equation}
where ${\chi\prime}^2$ is the reduced ${\chi\prime}^2$  , i.e. the ${\chi}^2$ per degree of freedom ($Df$). For this analysis, we define

\begin{equation}
{\chi\prime}^2=\frac{1}{Df} \sum_{i=1}^{N} w_i  \left(\frac{x^i_{model}-x^i_{obs}}{\sigma_i}\right)^2  \,, \label{eq15}
\end{equation}
where $N$ is the number of observed data points, while $x^i_{obs}$ and $x^i_{model}$ represent the $i$-th observed and modeled data point, respectively.  $\sigma^i$ is the error of the $i$-th observed data point. As the error of individual observations is unknown, we use the standard deviation of the observed $\phi_{total}$ and reconstructed $\phi_{open}$ data sets to estimate the values of ${\chi\prime}^2$. Following others \citep{holland1977}, we apply a weighting function (Cauchy weighting function) 

\begin{equation}
w = \frac{1}{1 + r^2} \,. \label{eq16}
\end{equation}

In order to reduce the influence of outliers, the value $r$ in the weight function is set to

\begin{equation}
r = \frac{\sqrt{h}} {\theta \left( \frac{mad(h)}{0.6745} \right)}  \,,\label{eq17}
\end{equation}
where $\theta$ is a tuning parameter, $h$ is the deviation of the model from the observations/reconstruction ($h=x^i_{model}-x^i_{obs}$). Here $mad$ is the median absolute deviation of the residuals from their median values, and the constant 0.6745 makes the estimate unbiased for the normal distribution. The tuning parameter ($\theta$) is set to 2.385 according to \cite{holland1977}. 

\subsection{Input data and parameters of the model}

For the total surface magnetic flux we use a set of observations compiled by \cite{arge2002} and \cite{wang2006}. This data set is based on almost daily observations of the solar global photospheric field that have been carried out at the Mt. Wilson Solar Observatory (MWO), National Solar Observatory Kitt Peak (KP NSO), and Wilcox Solar Observatory (WSO) over cycles 20-23. Note that for the optimization we only use data recorded between 1974 and 2001 when data from all observatories is available. Outside these periods we compare with the observations a posteriori as a further test of the model. We take into account the  finding of \cite{krivova2004} that more than half of the photospheric flux of ER may escape detection in the employed synoptic maps due to their relatively low spatial resolution.
Consequently, we compare the measurements of the total magnetic flux to the value ($\phi_{act}+c_{eph}\phi_{eph}+\phi_{open}$), where $c_{eph}$ is the fraction of $\phi_{eph}$ that is detected, which is relatively uncertain, because magnetic polarities are often missed on small scales. Here, we set $c_{eph}$ to be approximately 30\% compared to the value of 40\% employed by  \cite{krivova2007}. The value of $c_{eph}$ affects the amplitude of the ER cycle in the model since the values of the total flux during solar activity minima is determined mostly by the ER flux.

The parameters of the model as well as their adopted or best fit values are listed in Table \ref{table:1}. 
The ER flux decay time is fixed to a valued of 14 hours as found by \cite{hagenaar2001}. The remaining parameters are allowed to vary within a given range based also on independent observations and/or physical assumptions. This range is also given in Table \ref{table:1}.

\begin{table*}
\caption{Magnetic flux model parameters.}              
\label{table:1}      
\centering                                      
\begin{tabular}{l c r r r}          
\hline\hline
Parameter	& Symbol & Value (years)	& Min	& Max \\
\hline
AR Flux decay time scale	               & $\tau_{act}^0$  & 0.32   &  0.2	&    0.8    \\
AR Flux to Slow Open Flux transfer time scale  & $\tau_{act}^1$	     & 85.29  &  10.0   &   90.0    \\
AR Flux to Rapid Open Flux transfer time scale & $\tau_{act}^2$  & 1.71   &  0.0016	&    3.0016 \\
ER  Flux decay time scale	               & $\tau_{eph}^0$  & 0.0016 &  Fixed  &           \\
ER Flux to Slow Open Flux transfer time scale  & $\tau_{eph}^1$  & 10.08  &  10.0	&   90.0    \\
Rapid Open Flux decay time scale	       & $\tau_{open}^r$ & 0.1255 &  0.0822	&   0.3562  \\
Slow Open Flux decay time scale	               & $\tau_{open}^s$ & 1.36   &	 0.0016	&  6.0016   \\
ER amplitude factor	                       & $X$	     	 & 106.08 &  80.0   &  160.0    \\
ER cycle extension parameter	               & $c_x$	         & 5.01   &  5.0	&    9.0    \\
\hline
\end{tabular}
\end{table*}

The ranges of the decay time scale of the AR flux, $\tau_{act}^0$, the ER amplitude factor, $X$, and the extension parameter, $c_x$, are set as discussed previously by \cite{krivova2007} and \cite{solanki2000, solanki2002}. The best value found for the decay time scale of the AR flux is approximately 0.32 years. This estimate is close to the 0.25 years obtained assuming a balance between flux emergence and  decay \citep{krivova2007}. The best-fit ER amplitude factor is approximately 106 and the cycle extension parameter is approximately 5 years, which leads ER cycles that are longer than 20 years. These values are different from previouly obtained ones, which indicates that the introduction of $\phi_{open}^r$ changes the best-fit solution.

\cite{solanki2002} and \cite{krivova2007} postulates that the flux transfer time scale from ER to the open flux is a factor of six higher than the flux transfer time scale from AR ($\tau_{eph}^1=6\tau_{act}^1$). This assumption is based on the observation that in cycle 21 the contribution of ER to the axial dipole moment of the Sun was about a factor of six smaller than that of the AR, assuming an average lifetime of ER of 8 hours \citep{harvey1994}. Here, we have not constrained the value of $\tau_{eph}^1$ in relation to $\tau_{act}^1$ because it is not clear if only $\tau_{open}^s$ contributes to the axial dipole or if $\tau_{open}^r$ also influences it. In addition, we employ a longer average lifetime of ER (14 hours) as proposed by the more recent work of \cite{hagenaar2001}. We searched the best-fit values of both parameters in a wide range from 10 to 90 years. We defined this wide range in order search values in the domain in which \cite{krivova2007} found the solutions for this parameters. The best fit values for $\tau_{act}^1$ and $\tau_{eph}^1$ are approximately 85 and 10 years, respectively. 

Note that due to the extended length of the ephemeral cycle, around activity minimum both the preceding and following cycle contribute to the total and open flux. For the current minimum, we do not yet know the features (strength, length, and time of maximum) of the next cycle (cycle 24). 
Therefore we negled this cycle completely, so that the modeled magnetic flux values during the current minimum may be too low. 

We searched for the transfer time of AR flux to the rapidly decaying open flux ($\tau_{act}^2$) in the range between $\tau_{eph}^0$ (14 hours) and 6 years. The best-fit value found for this parameter is approximately 1.7 years, while the best value of the decay time of the rapidly evolving component of the open flux ($\tau_{open}^r$) is about 46 days. 

The decay time scale of the slowly evolving component of the open flux ($\tau_{open}^s$) is restricted to the range between 50 days and 6 years. The upper limit was defined in order to contain the values previously found for the decay time scale of the open flux (approximately 3-4 years). The optimum value returned by the code is approximately 1.4 years. This value is shorter than the previous estimate of the decay of the open flux between 3 and 4 years. 

We note that the set of best-fit parameters presented in Table 1 do not constitute a unique solution. In particular, we found solutions for short values of $\tau_{eph}^1$ (~3 years) and $\tau_{open}^s$ (~7 months) that have similar values of the fitness function. While a decay time of the open flux of about 3-4 years, as found by \cite{krivova2007, solanki2000, solanki2002}, determines the long-term evolution of the open flux by producing significant overlap between cycles, a short decay time of approximately 7 months cannot on its own account for the observed secular variations because the flux from the previus cycle has decayed before the next cycle starts properly. Instead, for such a short $\tau_{open}^s$ most of the long-term variability of the slowly evolving component of the open flux is due to the flux transferred from ER. For $\tau_{open}^s=1.4$ years both process play a role. 

\section{Results}

\subsection{Comparison of the model output and the observations/reconstructions}

\subsubsection{Total Magnetic Flux}

Figure \ref{FigTotalFlux}a shows a comparison of the observed (symbols) with the modeled (solid line) total magnetic flux. The three observational data sets are represented using different symbols: circles represent the KP NSO data, squares MWO data, and diamonds WSO data. Values are given for each Carrington rotation (CR) from the start of observations to the present:  CR 1615-1975 (NSO), CR 1516-2082 (MWO), and CR 1642-2081 (WSO). Only the period between the vertical dashed lines is used for the optimization of the parameters. The total flux plotted here is given by $\grave{\phi}_{total}= \phi_{act}+0.3\phi_{eph}+\phi_{open}$. The factor 0.3 takes into account that a major part of the photospheric flux from ER is missed when employing synoptic charts due to their relatively low spatial resolution \citep{krivova2004}. 

The model reproduces well the average variability of the data sets employed. The minima between the cycles 21-22 and 22-23 are slightly overestimated while the minima between the cycles 20-21 and 23-24 are well reproduced. Note that the observations of the descending phase of cycle 23 and the minimum between the cycles 23-24 were not employed for the optimization of the parameters of the model. It is grafifying to see that the model reproduces the total magnetic flux during the cycle 20 and during the declining phase of cycle 23, although these data points were not used to constrain the solution and the low value of the flux during the current minimum lies well outside the range of the previous minima. 

Figure 1b displays the calculated evolution of the AR magnetic flux, ER flux, open flux and the total flux. In this panel the total flux is given by the expression: ${\phi}_{total}= \phi_{act}+\phi_{eph}+\phi_{open}$. The modeled minimum values of the total flux between the cycles 20-21 is reduced by approximately 20\% relative to the minimum value between cycles 21-22 and it has the same level as the minimum between the cycles 22-23. The modeled value of the present minimum is approximately one third of that following cycle 22. For the model discussed here the ER flux has lower variation over the activity cycle than previously found by \cite{krivova2007} and \cite{solanki2002}. This is due to the very extended length of the ephemeral regions cycle, leading to a larger overlap between them. This small variation of the ephemeral region flux is in between the variations proposed by \cite{harvey1993} and \cite{hagenaar2001}. Shorter ephemeral region cycles, with less overlap, produce results that are more similar to those of \cite{harvey1993}, as modelled by \cite{solanki2002} and \cite{krivova2007}. Even longer ephemeral region cycles, with even more overlap, produce maxima in $\phi_{act}$ in agreement with the results of \cite{hagenaar2001}.

The relative contribution of AR, ER and open flux to the total flux is presented in Figure 1c. The average contribution of the AR flux to the total flux is approximately 50\% during the activity maxima while the contribution from ephemeral regions is about 40\%. During the minima, the ER flux contribution is about 80\% while the contribution from AR flux is approximately 10\%. The open flux represents a small fraction of the total flux (approximately 10\% during all phases of the cycle).

\begin{table*}
\caption{Magnetic flux model parameters.}              
\label{table:2}      
\centering                                      
\begin{tabular}{l c r r r}          
\hline\hline
Parameter	& Symbol & Value (years)	& Min	& Max \\
\hline
AR Flux decay time scale		  & $\tau_{act}$	&    0.22  & 0.2    &   0.8 \\
AR Flux to Open Flux transfer time scale  & $\tau_{ta}$	&   18.46  & 10.0   &  90.0 \\
ER  Flux decay time scale	          & $\tau_{eph}$	&   0.0016 & Fixed   &       \\
ER Flux to Open Flux transfer time scale  & $\tau_{te}$	&   110.76 & $6\tau_{ta}$ &  \\
Open Flux decay time scale		  & $\tau_{open}$	&    4.87  & 0.0016 &    6.0016 \\
ER amplitude factor			  & $X$	        &  159.05  &  80.0  &  160.0 \\
ER cycle extension parameter	          & $c_x$	&    6.45  &   5.0  &    9.0 \\
\hline
\end{tabular}
\end{table*}

\subsubsection{Open Flux}

\begin{figure*}[t]
\centering
\includegraphics[width=13cm,clip=true, viewport=0.1cm 0.1cm 19cm 27cm]{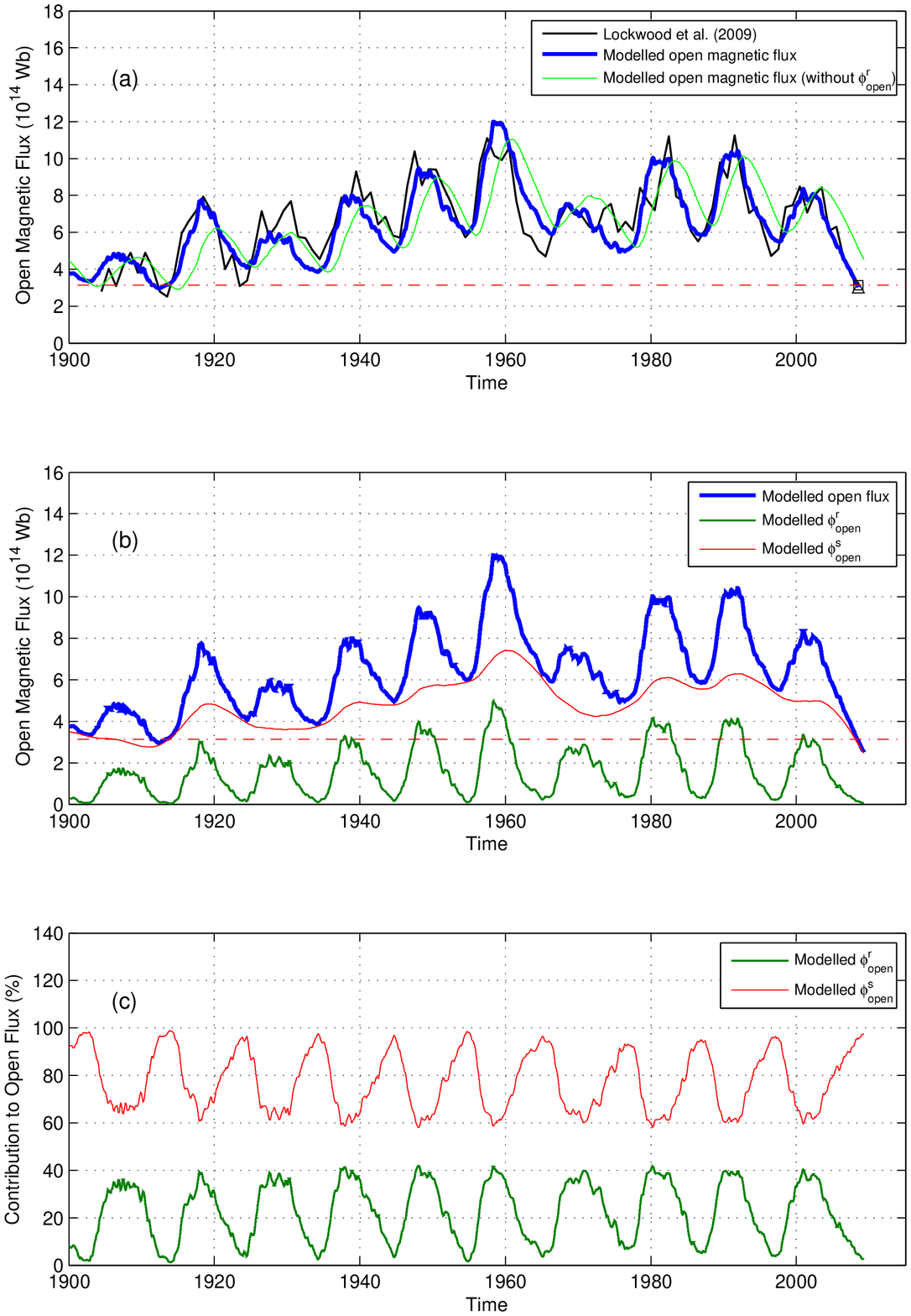}
\caption{(a) The open flux resulting from the present model (blue line) and the reconstruction based on the geomagnetic aa-index from 1904 to 2008 (black line). The dotted green line presents the open flux according to the model described by \cite{krivova2007}. The square marks the value observed in 2008, the triangle, the modeled value. Cycle number is indicated at the bottom of the panel. (b) Rapidly (green line) and slowly (red line) evolving components of the open flux (blue line). (c) Fractional contribution of the rapidly (green line) and slowly (red line) evolving components to the total open flux. For reference, dashed red lines representing the value observed in 2008 are drawn in panels (a) and (b).}
\label{FigOpenFlux}%
\end{figure*}

Figure \ref{FigOpenFlux}a shows a comparison between the open flux reconstructed by \cite{lockwood2009pers} (blackline) and the modeled open flux (blue line) from 1904 to 2008. The square indicates the value observed in 2008 (average over the year), the triangle the modeled value of the open flux (for the same period). For comparison, the green line represents the model described by \cite{krivova2007}. As the parameters of that model were obtained by fitting to the open flux reconstructed by \cite{lockwood1999}, we optimized the parameters of the model according to the procedure described in Sect. 2.3, i.e. to the data of \cite{lockwood2009pers}, prior to plotting its output in Fig. \ref{FigOpenFlux}a. The parameters obtained for the \cite{krivova2007} model are presented in Table \ref{table:2}, where the limits within which the parameters were searched for are also given. A lag of  approximately 2-3 years between the open flux reconstructed by \cite{lockwood2009pers} and that based on the \cite{krivova2007} model is evident. Phase shifts are also observed in flux transfer models due to the effect of the decay term in the flux transport equation \citep{mackay2002, schussler2006}. The calculated open flux evolution based on the present model reproduces well the reconstruction of the open flux based on the geomagnetic aa-index. In particular, the offset between model and observations, which is well seen in the results of \cite{solanki2002} and in the model based on \cite{krivova2007} is now gone. In addition, amplitudes of individual cycles are now in general better reproduced. For reference, a dashed red line representing the average value observed in 2008 is drawn. Clearly, the present model reproduces this value, while the previous, simpler version of the model employed by \cite{krivova2007} does not.

We note a discrepancy between the observed and modeled values for cycle 19, which is the strongest sunspot cycle observed since the Maunder Minimum. The cause of this discrepancy during the maximum of cycle 19 is not clear. It may be that for such a strong cycle, parameter values are different from those valid for other cycles. For example, \cite{wang2005} require different meridional flow speeds from cycle to cycle in order to reproduce the Sun's open magnetic flux and to obtain the polarity flip at the poles from one minimum to the next \citep{baumann2004}. The meridional flow speed could influence our parameters, which we have maintained unchanged for all cycles. Also, other solar parameters, such as the emergence latitudes,  could differs \citep{solanki2008}, which is not taken into account by our model. Alternatively, the reconstruction based on the aa-index could underestimate the peak value of the open flux during high solar activity. The sensitivity of high latitude stations to the auroral electrojet can be reduced at high activity because the electrojet drifts to lower latitudes \citep{lockwood2009c}. This effect is observed, for example, in the apparent saturation or even decrease of the geomagnetic AE-index during intense magnetic storms \citep{akasofu1981, feldstein1992,gonzalez1994}. As discussed by  \cite{lockwood2009c}, in principle, this non-linear effect is not significant for the aa-index because, although it is derived as a proxy of substorm activity, it is obtained from mid-latitude stations. Except during very intense magnetic activity, this effect is avoided for mid-latitude stations because the auroral electrojet always migrates toward the station with increasing activity \citep{lockwood2009c}. We speculate that due to the exceptionally high activity during cycle 19 the aa-index may have underestimate the level of maximum magnetic activity all the same.

According to \cite{wang2006}, the low-latitude component of the open flux closely tracks the Sun's equatorial dipole component, whose strength depends on the amount of flux present in the active regions and on the longitudinal distribution of the activity. Thus, asymmetries in the longitudinal distribution of large active regions can lead to large-amplitude variations of the low-latitude component of the open flux. As the model presented here does not describe the longitudinal distribution of the activity, the dips ("Gnevyshev gap") observed in the open flux near the cycle maxima are not reproduced. 

As pointed out in Sect. 2, in principle the magnetic flux emerging in ERs belonging to cycle 24 contributes to the slow component of the open flux during the descending phase of cycle 23 and the present minimum. However, the flux emerging in ER for cycle 24 was not included since we do not know the relevant cycle parameters yet. Nonetheless, the model reproduces the present low level of the open flux. 

This result suggests that cycle 24 will be rather weak or peak very late. However, since we cannot be sure that the deduced set of parameters represents a unique solution, we hesitate to use the present model to quantitatively predict the strength or length of cycle 24. Test calculations indicate that even for a given set value of the model's free parameters only a function $f(s,l)$ of the strength, $s$, and length, $l$, of the next cycle can be determined on the basis of our model. 

The evolution of the rapidly and slowly evolving components of the open flux is presented in Fig. 2b as well as the complete modeled open flux. Most of the cyclic variation is determined by the rapidly evolving component, while the slow component produces a background field that varies from cycle to cycle, but only relatively weakly over a cycle. The slow component peaks generally after the fast one in the descending phase of the sunspot cycle. The relative contributions of the rapidly and slowly evolving components to the complete open flux are shown in Figure 2c. A drop of similar magnitude as the current one (but starting from a higher level, so that the slow open flux did not reach such low levels as in 2008-2009) is seen between cycles 19 and 20, when a weak cycle followed a very strong one. During the solar minima, the open flux is maintained almost exclusively by the slow component (Figure 2c). According to the model, the rapidly evolving component contributes with about 40\% of the open flux at solar maxima and this fraction remained almost constant during the last century, even during cycle 19. 

\subsubsection{Estimates of errors in the model}

\begin{figure*}
\centering
\includegraphics[width=13cm,clip=true, viewport=0.1cm 0.1cm 19cm 18cm]{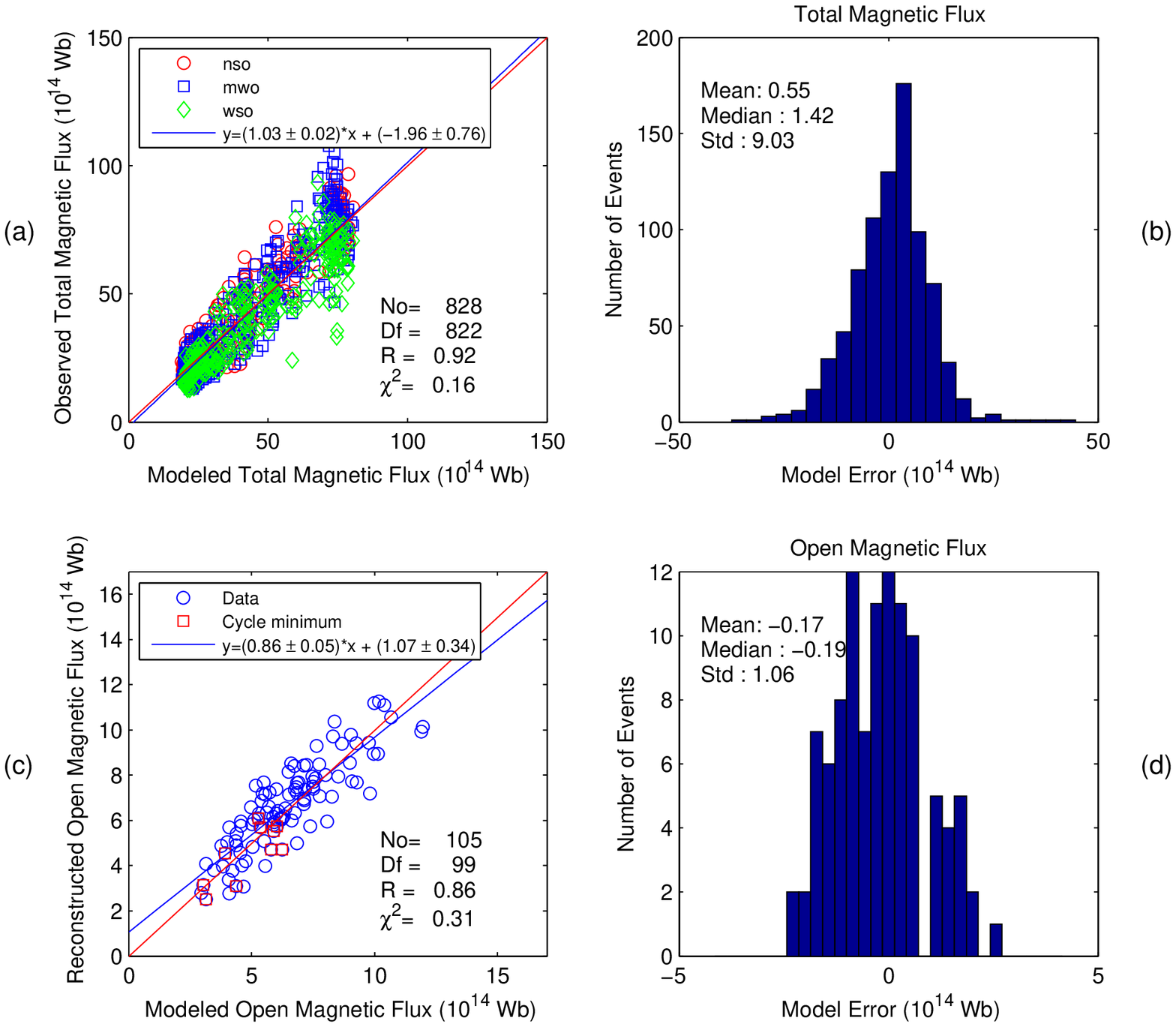}
\caption{(a) Scatter plot of the modeled versus the observed total flux. (b) Total Flux model error distribution. (c) Scatter plot of the modeled versus the reconstructed open flux. The cycle minima values (red squares) are indicated in the frame. (d) Open Flux model error distribution.}
\label{FigModelError}%
\end{figure*}

Figure \ref{FigModelError} presents further quantitative ways of comparing the model with the observations. Figure 3a displays the scatter plot of the modeled versus the observed total magnetic flux. As in Figure 1a, different data sets are indicated by different symbols. Each point represents a Carrington rotation. Also shown are a regression (blue line), whose equation is given in the panel, and the set of expectation values for the model ($y=x$; red line). The correlation coefficient for the overall data set is approximately 0.92 and the $\chi^2$ per degree of freedom (i.e. reduced $\chi^2$) is 0.16. Here, the reduced $\chi^2$ value is computed for the worst case with the weight function ($w$) in Eq. (15) equal to 1. The slope of the regression ($1.03 \pm 0.02$) indicates a good reconstruction of the variability of the cycle amplitude, although the amplitude of the reconstruction is somewhat lower than the observations. In Fig. 3b the distribution of the model error, defined here as the difference between the model output and the observations is plotted, for the total magnetic flux. The mean error value is $0.55\times10^{14}$ Wb and the standard deviation is about $9.03 \times10^{14}$ Wb. The distribution is slightly asymmetric with a median value equal to $1.42\times10^{14}$ Wb.  

The scatter plot of the modeled versus the empirically reconstructed open flux is plotted in Fig. 3c. Annual averages are plotted. The regression (blue line) and the set of expectation values for the model ($y=x$; red line) are also displayed. The correlation coefficient is approximately 0.86 and the reduced $\chi^2$, also computed with $w=1$, is 0.31. The slope of the regression is $0.86 \pm 0.05$. For this parameter, the regression is biased by the high difference between the values observed and modeled for the maximum of cycle 19 and ascending phase of cycle 21. The Figure 3d shows the distribution of the model error for the open flux. The mean and median values for the error distribution are $-0.17\times10^{14}$ Wb and $-0.19\times10^{14}$ Wb, respectively. The standard deviation is $1.06\times10^{14}$ Wb. 

\begin{figure*}
\centering
\includegraphics[width=13cm,clip=true, viewport=0.1cm 0.1cm 19cm 27cm]{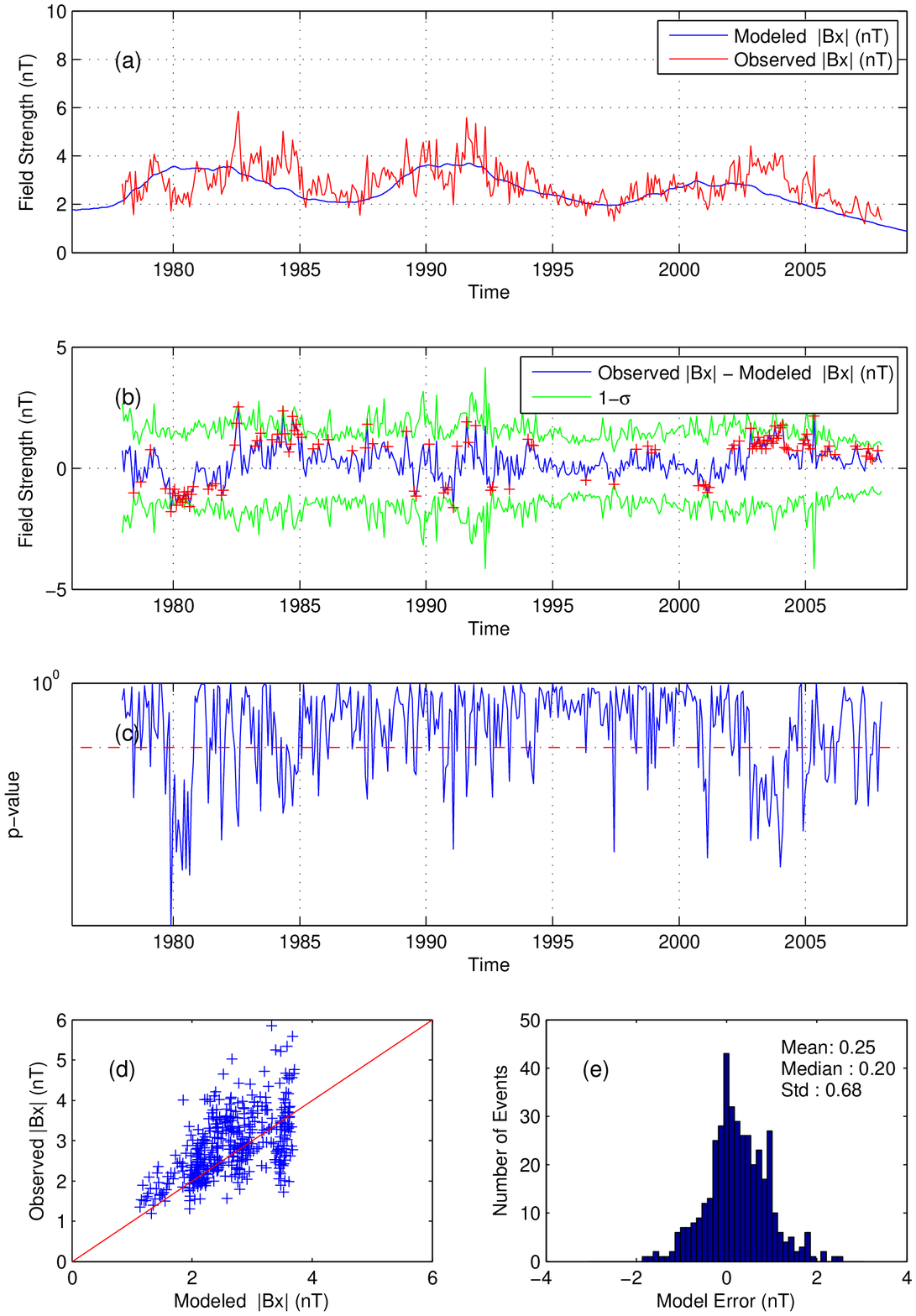}
\caption{(a) Modeled radial flux at 1 AU (blue line) in comparison with the measured radial interplanetary magnetic field component (OMNI data; red lines) averaged over Bartels rotations. (b) Difference between the measured and the modeled radial flux (blue line) at 1 AU. For reference, the 1-sigma value over Bartels rotations is also plotted (green lines). The read crosses (+)  represent the values rejected in the hypothesis test (see text for the description of the test). (c) Student's test p-values over Bartels rotations. The red dashed line shows $\alpha=0.05$. (d) Scatter plot of the modeled versus observed radial flux at 1 AU. (e) Distribution of the difference between the measured and the modeled radial flux  at 1 AU. }
\label{FigIMF}%
\end{figure*}

Since the open flux is distributed isotropically at 1 AU \citep[][and references therein]{lockwood2009a}, the radial field intensity at Earth is related to the open flux by

\begin{equation}
B_r (t)= \frac{\phi_{open}(t)}{4\pi r_E^2} \,, \label{eq18}
\end{equation}
where $r_E$ is the mean distance between the Earth and the Sun. The observed \citep[red line; OMNI 2 data,][]{king2005}{} and the modeled (blue line) radial component ($\left|Br\right|$) of the interplanetary magnetic field (IMF) are compared in Fig. \ref{FigIMF}a. The observations are averaged over Bartels rotations and cover the period from 1976 to 2009. 
Figure \ref{FigIMF}b presents the difference between the observations and the modeled radial component of the IMF (blue line) and the standard deviation of values within a Bartels rotation (dotted green lines). We note that the observed values are systematically higher than the modeled values. The mean difference is about 0.25 nT with a standard deviation of 0.68 nT (see Figs. 4d-e). In order to understand this discrepancy of 0.25 nT, we recall that to estimate the model parameters we compared the modeled output with the reconstruction by \cite{lockwood2009pers}, which incorporates a correction due to Kinematic effects \citep{lockwood2009a, lockwood2009b}. This correction effectively reduces the estimated open flux from interplanetary data. Consequently, the modeled radial component is systematically lower than the interplanetary observations, that do not incorporate this correction.

We test the null hypothesis that the mean value of $\left|Br\right|$ on each Bartels rotation is the one computed by the model. Here, we apply the Student's test. Figure 4c presents the p-values within Bartels rotations, which are the probability of observing a value as extreme or more extreme of the test statistic value given by

\begin{equation}
t_{test}= \frac{\left<x\right>-\mu}{\frac{s}{\sqrt{n}}}
\end{equation}
where $\left<x\right>$ is the mean, $\mu$ is the modeled value, $s$ is the sample standard deviation, and $n$ is the sample size. For reference, the red dashed line displays the 5\% significance level ($\alpha=0.05$). The red crosses in Fig. 4b indicate the periods at which we could reject the null hypothesis at the significance level of 5\%. We note several periods in which the model values do not represent the averaged values over Bartels rotations. Long periods of discrepancy occur in the ascending phase of cycle 21 and the descending phases of cycles 21 and 23. 


\subsection{Reconstruction of the solar magnetic fluxes since the Maunder Minimum}

\begin{figure*}
\centering
\includegraphics[width=13cm,clip=true, viewport=0.1cm 0.1cm 19cm 24cm]{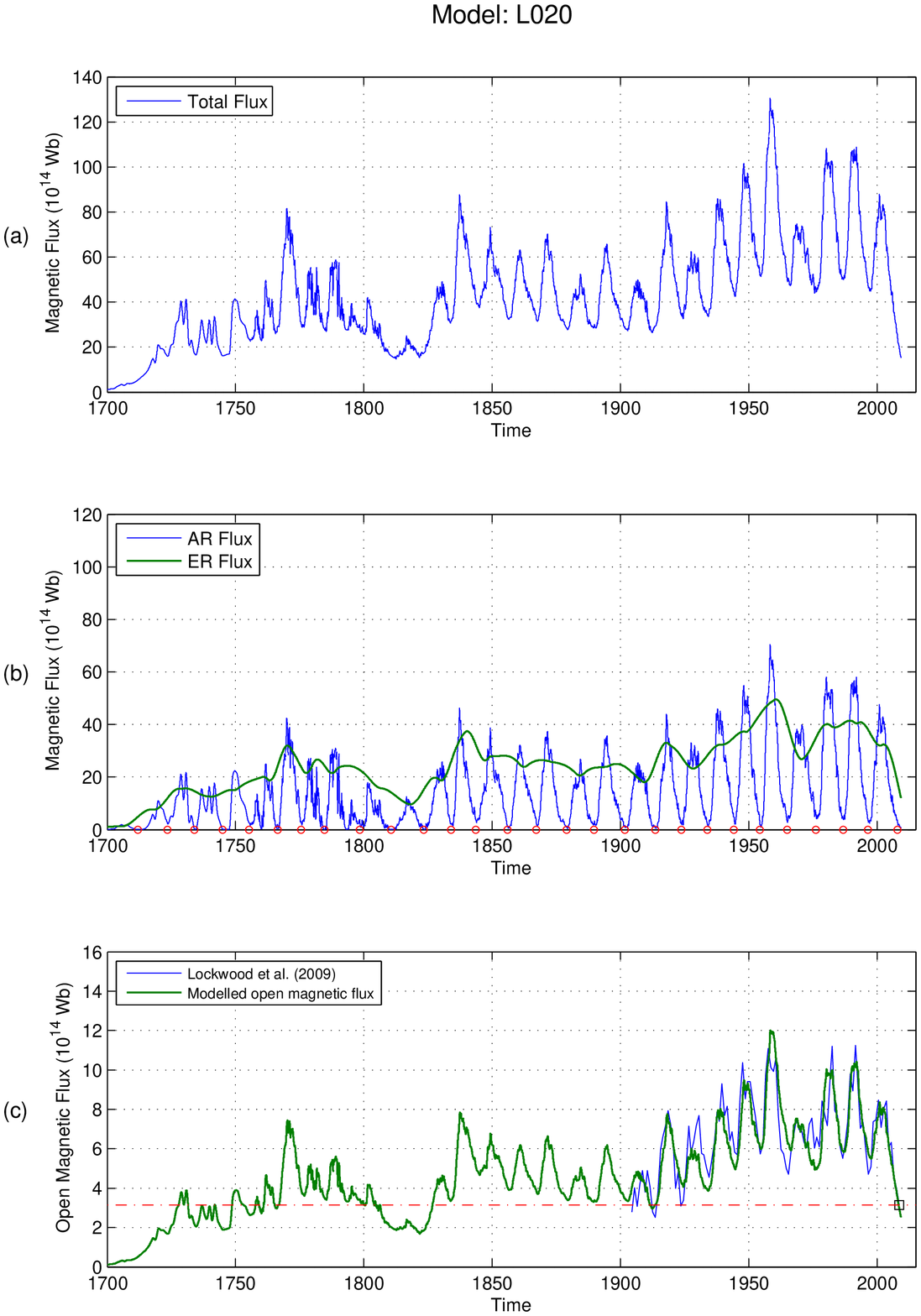}
\caption{Magnetic flux reconstruction from group sunspot number \citep{hoyt1998} since 1700. (a) Total Magnetic Flux. (b) Active (blue) and Ephemeral (green) Region fluxes. (c) Modeled open flux (green line). The Open Flux reconstruction based on the geomagnetic index-aa by \cite{lockwood2009pers} is plotted for reference (blue line). The dash red line represents the 2008 value of the Open Flux.}
\label{FigMagFluxMM}%
\end{figure*}

The reconstruction of the total flux ($\phi_{total} = \phi_{act} + \phi_{eph} + \phi_{open}$) from 1700 to 2008 based on the group sunspot number is displayed in Fig. \ref{FigMagFluxMM}a, while the reconstructions of the AR (blue line) and ER (green line) fluxes are plotted in Figure \ref{FigMagFluxMM}b. Finally, the reconstruction of the open flux (green line) is shown in Fig. \ref{FigMagFluxMM}c. For reference, the empirical reconstruction of the open flux (blue line) based on the geomagnetic aa-index is included in the plot. A clear secular trend in the ER and open flux leads to a secular trend in the total flux. 

As pointed out in Sect. 3.1.1, the modeled value of the total flux during the present minimum is approximately one third of the value observed during the previous minimum between cycles 22-23. Furthermore, the modeled total flux during the present minimum is at the same level of the value returned by the model for the Dalton Minimum, a period of low solar activity lasting from approximately 1790 to 1830 (Figure 5a). The modeled total flux during the present minimum is maintained entirely by the contribution of magnetic flux emerging in ERs belonging to cycle 23 since the flux emerging in ER for cycle 24 was not included due the lack of knowledge of parameters for this cycle. It must terefore be considered a lower limit.

The group sunspot number, which extends from approximately 1610 to the present, allows the solar magnetic flux to be reconstructed also prior to 1700. The model gives an open and total flux near zero through most of the Maunder Minimum (MM). This is a natural consequence of the coupling of the strength of the cycle of ER to that of sunspots, so that $\phi_{eph}$ turns out to be extremely weak in the Maunder Minimum. In addition, the length of the Maunder Minimum is longer than the decay time of the flux, so that in the model practically no flux survives from the pre-Maunder Minimum cycles until the end of the MM.

\section{Reconstruction of solar magnetic flux for the Holocene}

For studies aiming to isolate the Sun's influence on the Earth's climate time series of the solar magnetic and activity are needed as long as possible. Unfortunately, indices of solar activity such as the sunspot number, used here to reconstruct the AR, ER,  total and open magnetic flux, have been adequately recorded only since the invention of the telescope in the 17th century. They represent the longest running time series of direct measurements of past solar variability. Cosmogenic isotopes provide estimates of solar activity that are less clean, in the sense that they are affected by the other quantities, such as the geomagnetic field and processes (e.g. climate variations and carbon cycle). However, these records extend to earlier times, covering periods up to thousands of years \citep{stuiver1989}. 

By combining physics-based models for each of the processes connecting the isotope concentration in a relevant terrestrial archive with solar activity, open flux and from it the sunspot number could be reconstructed by \cite{solanki2004} and \cite{usoskin2004, usoskin2007}. Since from cosmogenic isotope records the primary solar parameter that can be determined is the open flux, the problem faced here is opposite the one dealt with in Sect. 2 and 3. How to compute the SN from the open flux? This was first dealt with by \cite{usoskin2002}. In the reconstructions carried out so far, the solar open flux is linked with the sunspot number, AR and ER flux by inverting the model by describing the evolution of the solar surface magnetic components for a given sunspot number \citep{krivova2007, solanki2000, solanki2002}. 
In this section, we update the previous estimate of the sunspot number, AR, and ER flux based on the inversion of the model described in Sect. 2.1. In addition, often only multi-year, e.g. decadal, averaged data are available. Hence we need to take this into account. 

\subsection{Derivation of the magnetic flux model}

By differentiating Equation (\ref{eq06}) with respect to time, we obtain
\begin{equation}
\frac{d\phi_{open}}{dt}=\frac{d\phi_{open}^r}{dt}+\frac{d\phi_{open}^s}{dt} \,. \label{eq20}
\end{equation}

Substituting Equations (\ref{eq04}) and (\ref{eq05}) in (\ref{eq20}), we find
\begin{equation}
\frac{d\phi_{open}}{dt} = \frac{d\phi_{open}^r}{dt} + \frac{\phi_{act}}{\tau_{act}^1} + \frac{\phi_{eph}}{\tau_{eph}^1} + \frac{\phi_{open}^r}{\tau_{open}^s} - \frac{\phi_{open}}{\tau_{open}^s} \,. \label{eq21}
\end{equation}

After averaging over 10 years, we obtain
\begin{equation}
\label{eq22}
\left<\frac{d\phi_{open}}{dt}\right> = \left< \frac{d\phi_{open}^r}{dt} \right> + \left< \frac{\phi_{act}}{\tau_{act}^1} \right> + \left< \frac{\phi_{eph}}{\tau_{eph}^1} \right> + \left< \frac{\phi_{open}^r}{\tau_{open}^s} \right> - \left< \frac{\phi_{open}}{\tau_{open}^s} \right> \,. 
\end{equation}
Here, the symbol $\left< ... \right>$ denotes 10-year averaging. 

We assume that on decadal time scale the flux in active regions evolves in a steady state, i.e. the flux emerging in active regions is approximately equal to the decay due to several processes. In this way, we can write from Eq. \ref{eq01}

\begin{equation}
\label{eq23}
\left< \frac{\phi_{act}}{\tau_{act}} \right> = \left< \epsilon_{act} \right> \,,
\end{equation}
where
\begin{equation}
\label{eq24}
\frac{1}{\tau_{act}} = \frac{1}{\tau_{act}^0} + \frac{1}{\tau_{act}^1} + \frac{1}{\tau_{act}^2} \,. 
\end{equation}
The validity of this approximation can be tested by computing the ratio $\left< {\phi_{act}} \right> / \left< \epsilon_{act} \right>$ from the model output, which should be close to $\tau_{act}$ (~0.27 years). We found that $\left< \phi_{act} \right> / \left< \epsilon_{act} \right>$ is approximately 0.1\% higher than the value of $\tau_{act}$. Consequently, the assumption that the flux in active regions evolves in a steady state on a decadal scale is well founded. Substituting Eq. (\ref{eq07}) into Eq. (\ref{eq23}), we obtain

\begin{equation}
\left< \phi_{act} \right> \approx \tau_{act} \frac{\epsilon_{act}^{max,21}}{R_g^{max,21}} \left< R_g \right> \,. \label{eq28}
\end{equation}

Similarly, since $\tau_{eph}^0 \approx 14$ hours is also very short compared to the cycle length, we obtain from Eqs. (\ref{eq02}) and (\ref{eq08}) 

\begin{equation}
\phi_{eph}(t) \approx \tau_{eph} \epsilon_{eph}(t) \approx \tau_{eph} X \sum_{i=1}^{N_{cycles}} \epsilon_{act}^{max,i} g^i(t) \,,\label{eq30}
\end{equation}
where

\begin{equation}
\frac{1}{\tau_{eph}} = \frac{1}{\tau_{eph}^0} + \frac{1}{\tau_{eph}^1} \,.\label{eq31}
\end{equation}

The maximum emergence rate during the activity cycle $i$ can be computed using Equation (\ref{eq13}). 
Following \cite{usoskin2007}, we assume a linear relation between the amplitude of the solar cycle and the 10-year averaged sunspot number ($R_g^{max,i}=k·<R_g>$. The numerical value of k found by  \cite{usoskin2007} is $2.2 \pm 0.4$. The flux emergence in ER displays extended cycles, so that adjacent cycles partially overlap. For the set of parameters presented in Table 1, adjacent cycles overlap in a way that a low 11-year variability of the ER cycle is observed and the long-term trend is directly related to the 10-year averages of $R_g$. At the maximum of an ephemeral region cycle, $\phi_{eph}$, we can set $g_i=1$, so that we obtain using Eq. (\ref{eq13})

\begin{equation}
\label{eq32}
\left< \epsilon_{act}^{max,i} \right> \approx \frac{\epsilon_{act}^{max,21}}{R_g^{max,21}} R_g^{max,i} \approx \frac{\epsilon_{act}^{max,21}}{R_g^{max,21}} k \left< R_g \right> \,.
\end{equation}

The ER flux averaged on a decadal time scale is then given by
\begin{equation}
\label{eq33}
\left< \phi_{eph} \right> \approx \tau_{eph} \frac{\epsilon_{act}^{max,21}}{R_g^{max,21}} k X \left< R_g \right>  \,.
\end{equation}

We can also assume that the rapid open flux evolves in a steady state on a decadal time scale. In this case, 
\begin{equation}
\frac{<\phi_{act}>}{<\phi_{open}^r>} = \frac{\tau_{open}^r}{\tau_{act}^2} \,.
\end{equation}
We found that the value of the ratio ${<\phi_{act}>}/{<\phi_{open}^r>}$ is within approximately 0.5\% of the value of ratio ${\tau_{open}^r}/{\tau_{act}^2}$. Consequently, for a steady state, the term $ \left< \frac{d\phi_{open}^r}{dt} \right>$ in Eq. (\ref{eq22}) can be negleted.

In order to compute the evolution of the open flux, we assume that 

\begin{equation}
\label{eq35}
\left< \frac{d\phi_{open}}{dt} \right> \approx \frac{\Delta\left<\phi_{open}\right>}{\Delta t} \,,
\end{equation}
where 

\begin{equation}
\label{eq36}
\left<\Delta\phi_{open}\right> = \left<\phi_{open}(t_j+\Delta t)\right>- \left<\phi_{open}(t_j )\right> \,.
\end{equation}

Substituting Eqs. (30) and (31) into (\ref{eq22}) we obtain

\begin{equation}
\label{eq37}
\frac{\left< \phi_{open} \right>_{j+1}}{\Delta t} + \frac{\left< \phi_{open} \right>_j}{\tau_1} = 
\left( \frac{1}{\tau_{act}^1} + \frac{\tau_{open}^f}{\tau_{open}^s \tau_{act}^2}  \right) 
\left< \phi_{act} \right>_j + 
\frac{\left< \phi_{eph} \right>_j}{\tau_{eph}^1} \,,
\end{equation}  
where $\left< \phi_{open} \right>_{j+1} = \left<\phi_{open}(t_j+ \Delta t) \right>$, $\left< \phi_{open} \right>_j = \left< \phi_{open}(t_j) \right>$, and 

\begin{equation}
\label{eq38}
\frac{1}{\tau_1} = \frac{1}{\tau_{open}^s} - \frac{1}{\Delta t} \,.
\end{equation}

Replacing Eqs. (25) and (29) into (33) we obtain

\begin{equation}
\label{eq39}
\frac{\left< \phi_{open} \right>_{j+1}}{\Delta t} + \frac{\left< \phi_{open} \right>_j}{\tau_1} =
c \left< R_g \right>_j \,,
\end{equation}
where, the constant c is given by

\begin{equation}
\label{eq40}
c = \left[ \left( \frac{1}{\tau_{act}^1} + \frac{\tau_{open}^r}{\tau_{open}^s \tau_{act}^2}\right) \tau_{act} + \frac{\tau_{eph}kX}{\tau_{eph}^1} \right] \frac{\epsilon_{act}^{max,21}}{R_g^{max,21}} \,.
\end{equation}

Rearranging, we get

\begin{equation}
\label{eq41}
\left< R_g \right>_j=a \left< \phi_{open} \right>_j + b \left< \phi_{open} \right>_{j+1} \,,
\end{equation}
where

\begin{equation}
\label{eq42}
a = \frac{1}{c \tau_1} \,,
\end{equation}
and

\begin{equation}
\label{eq43}
b = \frac{1}{c \Delta t} \,.
\end{equation}

The 10-year averaged AR and ER can be retrieved by substituting Eq. (37) in Eqs. (25) and (29), respectively. The rapidly evolving component of the open flux can then be obtained from Eq. (30).

\begin{figure}
\centering
\resizebox{\hsize}{!}{\includegraphics[clip=true, viewport=0.1cm 0.1cm 17cm 12cm]{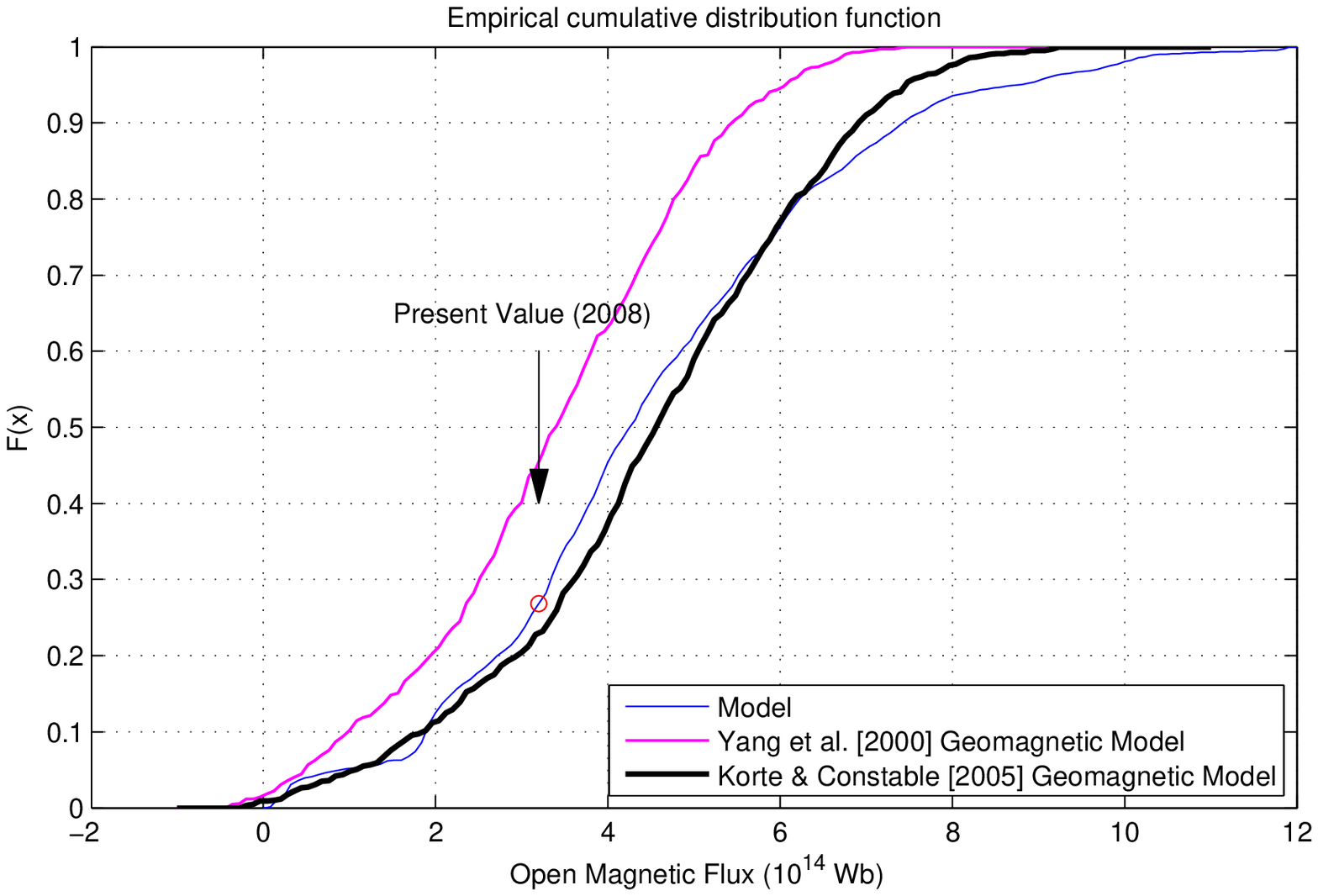}}
\caption{Comparison of the empirical cumulative distribution functions, F(x), of the open flux. The blue line is the distribution of the modeled open flux based group sunspot number \citep{hoyt1998}. Red and black lines are distributions of the open flux based on \element[ ][14]{C} data \citep{usoskin2007} derived from paleo-geomagnetic reconstructions of \cite{yang2000} and \cite{korte2005}, respectively. The 2008 annual average is indicated by the red circle. }
\label{FigDistribution}%
\end{figure}

\begin{figure*}
\centering
\includegraphics[width=13cm,clip=true, viewport=0.1cm 0.1cm 19cm 18cm]{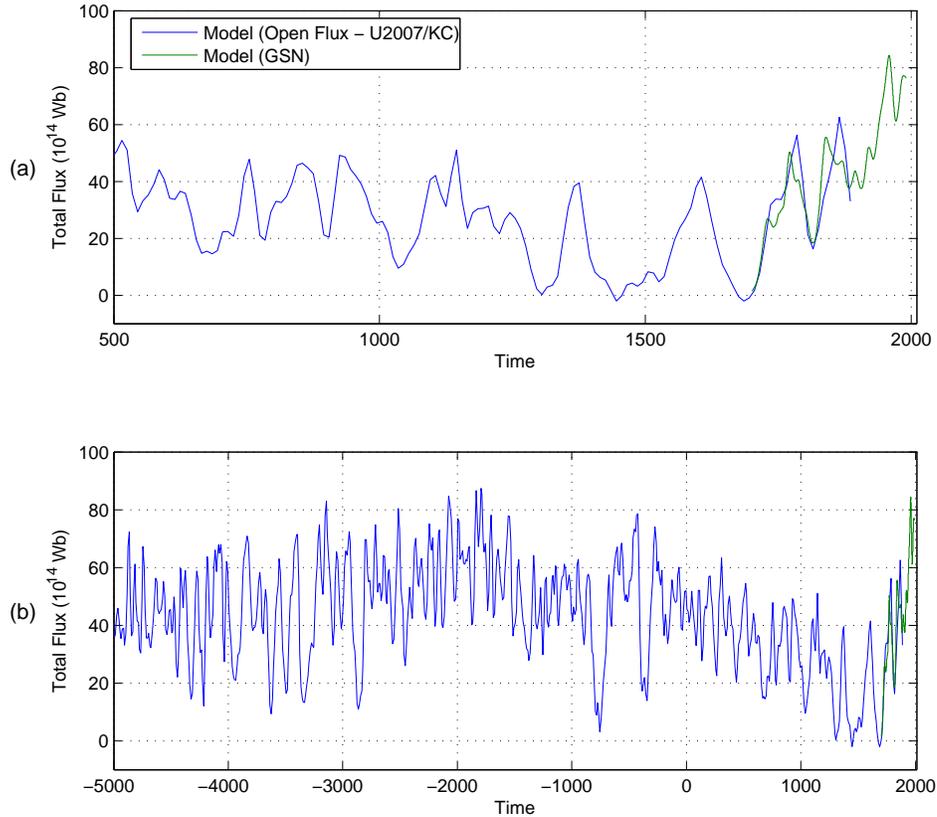}
\caption{Long-term total magnetic flux reconstruction from 14C data. The blue line is the reconstruction based on the open flux obtained by \cite{usoskin2007}using paleo-geomagnetic data from \cite{korte2005}. The green line is the reconstruction of the total magnetic flux (10-year running means) based on the group sunspot number since 1700 AD. Panel (a) shows the reconstruction from 500 AD to present while panel (b) shows the reconstruction since 5000 BC.}
\label{FigTotalFluxHolocene}
\end{figure*}


\subsection{Reconstructed magnetic flux through the Holocene}

Our reconstruction of past solar activity relies on the decadal estimate of the solar open flux from measurements of \element[ ][14]{C} as was done earlier by \cite{solanki2004} and \cite{usoskin2007}. As discussed by \cite{usoskin2007}, the estimate of the open flux depends on the knowledge of the temporal evolution of the geomagnetic field. \cite{usoskin2007} presented two reconstructions based on the paleomagnetic models by \cite{yang2000} and \cite{korte2005}. The first one extends through the whole Holocene while the second one reaches back around 7000 years. In Fig. \ref{FigDistribution}, a comparison between the empirical cumulative distribution functions of these two reconstructions based on \element[ ][14]{C} data and the one based on the telescopic sunspot record is presented. In order to compare the distributions of the open flux based on \element[ ][14]{C}, we have employed just the period over which the time series overlap. As the geomagnetic dipole moment of \cite{korte2005} is systematically lower than that obtained by \cite{yang2000}, a systematically higher open flux is obtained. We note that the distribution of the open flux estimated from sunspot number is closer to the reconstruction using the \cite{korte2005} geomagnetic dipole moment. It suggests that if the paleo-geomagnetic reconstructions of \cite{korte2005} is close to the real evolution of the magnetic field, the values of the  open flux since Maunder Minimum are not unusual comparing to the values observed during the Holocene. If, however, the paleo-geomagnetic reconstruction by \cite{yang2000} is closer to the real evolution of the magnetic field, the values of the open flux since the Maunder Minimum are unusually high compared to the Holocene. Furthermore, we note that about 25\% of the modeled values of the open flux from these two reconstructions are below the value observed in the present minimum based on the  \cite{korte2005} reconstruction, while about 45\% of the values of the open flux based on \cite{yang2000} are below the present value. Note that the value for the present minimum is a yearly value, while the curves based on the reconstructions from \element[ ][14]{C} are decadal averages.

A comparison between the reconstructions of the sunspot number obtained using Equation (37)  and the one obtained by \cite{usoskin2007} reveals a good correspondence ($R \approx 0.96$). We note that the reconstructions obtained based on the two approaches are quite similar, with the reconstruction from \cite{usoskin2007} having slightly lower values during high activity. Both reconstructions are based on the open flux estimate by \cite{usoskin2007} using the \cite{korte2005} model. The main difference between the two models is the relationship between the open flux and sunspot number. In this work, we distinguished the fast and slowly evolving components of the open flux, while in the reconstruction by \cite{usoskin2007} such a separation is not made.

We obtain the reconstruction of the total magnetic flux from the estimate of the magnetic flux in AR (Eq. (25)), ER (Eq. (29)), and the open flux provided by \cite{usoskin2007}. Figure \ref{FigTotalFluxHolocene} presents the obtained reconstruction based on the \cite{korte2005} geomagnetic model. In Fig. 7a, we concentrate on the reconstruction since 500 AD (blue line). For reference, the total flux estimated from sunspot data is presented. The reconstruction since 5000 BC is shown in Figure 7b. The total flux is required to compute the irradiance, which will be the topic of a forthcoming paper.

\section{Concluding remarks}

In the present paper, we have considered an extension of the simple model of \cite{solanki2002} and \cite{krivova2007} describing the evolution of the Sun's open and total magnetic flux. We have shown that by considering separately a rapidly evolving and a slowly evolving component of open flux we obtain a greatly improved agreement with of the solar open flux reconstructed since 1904 by \cite{lockwood2009pers} and a reasonable although not perfect description of the OMNI data as well as. The main improvement provided by this version of the model is in the reproduction of the cyclic variation of the open flux, including the amplitudes of individual cycles. 

The rapidly decaying open flux is most likely harbored in small coronal holes associated with ARs or decaying ARs, while the slowly decaying open flux is associated with the polar coronal holes, but also with open flux at low latitudes. 

We found that approximately 25\% of the modeled open flux values since the end of the Maunder Minimum are lower than the low observed during the present minimum (i.e. in 2008), which suggests that the present solar minimum conditions are not exceptional in terms of the heliospheric magnetic flux. We noted that the same amount is observed in the reconstruction of the open flux by \cite{usoskin2007} based on the \cite{korte2005} model, while about 45\% of the open flux values are lower than the value during the current minimum in the reconstruction based on the \cite{yang2000} model.

\begin{acknowledgements}
 We would like to thanks N. Krivova for the helpful and fruitful discussions. We thank C.N. Arge, J.W. Harvey, H.P. Jones, Y.-M. Wang and N.R. Sheeley Jr. for providing the photospheric magnetic flux data, as well as M. Lockwood for providing the open flux reconstructions based on the aa-index. The PIKAIA optimization subroutine used in this work is available from http://download.hao.ucar.edu/archive/pikaia/. The OMNI data were obtained from the GSFC/SPDF OMNIWeb interface at http://omniweb.gsfc.nasa.gov. The sunspot data were obtained from the National Geophysical Data Center (NGDC, http://www.ngdc.noaa.gov/stp/SOLAR/solar.html). This work has been partially supported by the WCU grant No. R31-0016 funded by the Korean Ministry of Education, Science and Technology.
\end{acknowledgements}

\bibliography{13276ref}{}
\bibliographystyle{aa} 

\end{document}